\documentclass{article}%
\usepackage{amssymb}%
\usepackage{amsmath}%
\usepackage{amsfonts}%
\usepackage{graphicx}

\begin{document}

\begin{center}
{\Large Thermodynamic basis of the concept of "recombination resistances"}%
\[
\]

M. Salis(*), P. C. Ricci and F. Raga

\textit{Dipartimento di Fisica - Universit\`{a} di Cagliari}

\textit{INFM- UdR CagliariCittadella Universitaria , 09042 Monserrato-Cagliari
, Italy}
\end{center}

The concept of \textquotedblright recombination resistance\textquotedblright%
\ introduced by Shockley and Read (Phys. Rev. \textbf{87}, 835 (1952)) is
discussed within the framework of the thermodynamics of irreversible processes
ruled by the principle of the minimum rate of entropy production. \ It is
shown that the affinities of recombination processes represent "voltages" \ in
a thermodynamic Ohm-like law where the net rates of recombinations represent
the "currents". The quantities thus found allow for the definition of the
\textquotedblright dissipated power\textquotedblright\ which is to be related
to the rate of entropy production of the recombination processes dealt with.


\[
\]

\begin{center}
{\large I.} {\large INTRODUCTION}
\end{center}

Crystal defects are physical entities of \ primary importance in solid state
physics. Owing to the breaking of translational symmetry, they originate
localized levels capable of trapping electrons or holes. On this ground, the
physical properties of materials such as semiconductors can be modified by
suitable doping determining the sign of majority carriers. Impurity or
intrinsic defects may originate localized levels allowing recombinations of
electron-hole pairs injected into bands by some excitation mechanism, thus
affecting the lifetime of free carriers. The content of intrinsic defects at
(lattice) equilibrium is ruled by general laws$^{1}$. But in reality it
strongly depends on the history of the sample dealt with. Depending on the
nature of the defects, three basic schemes are to be considered. They
are:\ the Shon Klasens (SK) scheme, in which a conduction band electron
recombines with a hole kept in a localized level, the opposite Lamb-Klick (LK)
scheme, in which a valence band hole recombines with a localized electron, and
the Prener Williams scheme, in which both electron and hole are localized
within a two level center$^{2}$. \ 

At equilibrium, \ the \ occupation \ of electron (hole) levels is ruled by
\ the \ Fermi-Dirac \ distribution \ \ $f_{n}(E)$ ($f_{p}(E)=1-f_{n}(E)$),
that is, $f_{n}(E)=$ $1/\left\{  1+\exp\left[  \left(  E-E_{F}\right)
/kT\right]  \right\}  $ where $E$ and $E_{F}$ stand for the actual and the
Fermi levels. The equilibrium densities of conduction band electrons and
valence band holes are$^{3}$ \ $\overline{n}=N_{n}\exp\left[  \left(
E_{F}-E_{C}\right)  /kT\right]  $ and $\overline{p}=N_{p}\exp\left[  \left(
E_{V}-E_{F}\right)  /kT\right]  $, respectively, where $E_{C}$ and $E_{V}$ are
the energy levels of the bottom of the conduction band and of the top of the
valence band, respectively, $N_{n}=2\left(  2\pi m_{e}kT/h^{2}\right)  ^{3/2}$
\ and $\ N_{p}=2\left(  2\pi m_{p}kT/h^{2}\right)  ^{3/2}$ , $m_{e}$ and
$m_{p} $ the electron and hole effective masses, respectively . \ It is to be
pointed out that distinction between electron and hole traps is considered for
convenience since only the capture cross section determines the trapping
properties. \ Often, in the physics of semiconductors a hole trap is presented
as a deep electron level, while a hole trapping level is presented as an
acceptor level. In the physics of ionic crystals it is usual to define the
trapping properties by means of the defect-charge states. Thus, negative
charge-defects are traps for holes (positive carriers). On the contrary,
positive-charge defects are traps for \ electrons (negative carriers). But
there are cases where these rules do not hold ( for example, the U$^{-}$
centres)$^{4}$. Also the distinction of carrier capture for trapping or
recombination may be considered as conventional, since in both cases we are
dealing with changes in level occupancies. Thus, \ for a general discussion
about capture processes it is convenient to follow the model of trap
classifications used by Simmons and Taylor (ST)$^{5}$: \textit{"when a trap is
empty it is ready to receive an electron, \ and thus it is operating as an
electron trap. When the trap contains an electron, it is ready to receive a
hole, and hence is a hole trap. (We are assuming that the traps are
monovalent.) It is convenient to assume that the traps existing below \ the
equilibrium Fermi level are neutral when filled with an electron and that the
traps positioned above the equilibrium Fermi level are neutral when empty.
(...) Thus a trap positioned above the equilibrium Fermi level is neutral when
acting as an electron trap and negatively charged when acting as a hole trap.
\ On the other hand, a trap positioned below the equilibrium Fermi levels is
neutral when acting as a hole trap and positively charged when acting as an
electron trap."}

\ Departure from equilibrium can be obtained by different excitation sources.
In this paper, we are considering ionizing photons causing band-to-band
transitions. As the excitation source is turned on, carriers injected into the
conduction and valence bands are drawn by the several processes occurring in
the material, including trapping in metastable levels. The latter process
allows crystals to attain excited states which can be held after the
excitation source is turned off. In this case, the fundamental state can be
reached by thermally stimulated processes$^{6}$. Under steady excitation, the
population of electron- or hole-levels attain a steady distribution after a
time which, in some cases, may be very long$^{7}$. The statistic of occupancy
of the traps may be obtained from two different points of view. The first of
these, which was used by Shokley and Read (SR)$^{8}$, considers the rate
equations \ for the conduction and valence bands. The second, which was used
by ST$^{5}$ , considers the rate equations for a particular trapping centre.
In the non-degenerate case, both these approaches lead to an occupation $f(E)$
of trap levels with energy $E$ and density $N_{t}(E)$ which is given by%

\begin{equation}
f(E)=\frac{c_{n}n+c_{p}N_{p}\exp\left[  \left(  E_{V}-E\right)  /kT\right]
}{c_{n}\left\{  n+N_{n}\exp\left[  \left(  E-E_{C}\right)  /kT\right]
\right\}  +c_{p}\left\{  p+N_{p}\exp\left[  \left(  E_{V}-E\right)
/kT\right]  \right\}  }\label{INtro1}%
\end{equation}
where $c_{n}$ and $c_{p}$ stand for the electron and hole capture
probabilities, respectively. The basic assumption of the two approaches is
that, at the steady state, the population of electron- or hole-levels can be
described by means of a Fermi-Dirac-like function with suitable quasi-Fermi
levels (QFL). \ For free carriers it was found that%

\begin{equation}
\frac{np}{\overline{n}\overline{p}}=\exp\left[  (F_{n}-F_{p})/kT\right]
\label{IntroZZ1}%
\end{equation}
where $F_{n}$ and $F_{p}$ stand for QFLs of free electrons and holes, respectively.

QFLs play a role in the rate of recombination processes. A suggestive idea
introduced by SR considers the recombination rate at the steady state as a
current passing through a resistance , called a \textquotedblright
recombination resistance\textquotedblright\ (RR), depending on the kinetic
parameters of the process dealt with. In the simplest cases, recombinations
follow from two capture processes, that is, hole and electron capture, so that
two resistances are to be considered , that is, near equilibrium%
\begin{equation}
R_{n}=kT/\overline{n}\overline{p}_{t}c_{n}\label{IntroX1}%
\end{equation}
for electron capture and%

\begin{equation}
R_{p}=kT/\overline{p}\overline{n}_{t}c_{p}\label{IntroX2}%
\end{equation}
for hole capture, $\ \overline{n}_{t}=f_{t}N_{t}$ and $\overline{p}%
_{t}=\left(  1-f_{t}\right)  N_{t}$ \ standing for the equilibrium densities
of trapped electrons and holes, respectively, and $f_{t}$ \ for the level
occupancy as given by the Fermi-Dirac function. By presenting QFLs as "
voltages", SR proved that in the near-equilibrium approximation the above
resistance definitions allow for the equation%

\begin{equation}
\upsilon\left(  R_{n}+R_{p}\right)  =F_{n}-F_{p}\label{IntroX3}%
\end{equation}
which has the formal structure of Ohm's law. SR also derived an equation for
the lifetime of free carriers, that is, $\tau=\overline{n}\overline
{p}R/kT\left(  \overline{n}+\overline{p}\right)  $ where $R=R_{n}+R_{p}$.
\ About this point they concluded that \textquotedblright\textit{the effect of
a number of different sorts of traps may be considered on the same basis. For
each variety, the recombination is represented by a pair of resistances in
series and these series pairs are combined in parallel for the entire
system}\textquotedblright.

At the level of the SR treatment we cannot immediately give a thermodynamic
meaning to the product $\upsilon R^{2}$ \ which, in the theory of electricity,
corresponds to the dissipated electrical power. \ In all probability, for this
reason the concept of RR has remained an unexplored minor outcome of the SR
statistics although sometimes it is recalled in papers devoted to studies on
the kinetics of electronic processes in semiconductor based devices$^{9}$. The
scope of this paper is thus to investigate the recombination processes at the
steady state from the general point of view of the thermodynamics of
non-equilibrium. To this purpose, we remain in the range of linear
irreversible processes, where it is possible to apply successfully the
principle of minimum rate of entropy production (MREP). This approach
resembles the one used (for a different scope) by V. Maxia$^{10}$ . The goal
of this paper is to give a thermodynamic basis to the concept of RR. It is
shown that affinities are more suitable to represent \textquotedblright
voltages \textquotedblright\ than QFLs. What is better, it is shown that the
analogue of "dissipated power " is closely related to the rate of entropy
production of processes dealt with. Thus, the paper is structured as follows.
In Section II, basic concepts of non-equilibrium thermodynamics as well as the
MREP principle are recalled. A suitable variational calculus based on the MREP
principle is presented. \ In Section III, the results thus obtained are
applied to typical recombination processes. \ Final remarks are given in
Section IV.\
\[
\]

\begin{center}
{\large II. THERMODYNAMICS OF THE NEAR-EQUILIBRIUM STEADY STATE}%
\[
\]
{\large \ }

\textbf{A. Theory}
\end{center}

Non-equilibrium thermodynamics lacks \ a constructive criterion such as the
one given by maximum entropy for the equilibrium state. The latter provides a
starting point for the application of statistical mechanics and
thermodynamics. \ However, non-equilibrium thermodynamics shows that states
having minimum entropy production compatible with the system constraints are
stationary states$^{11}$. In reality, this property characterizes the case of
linear flux laws with constant phenomenological coefficients. Nevertheless,
MREP allows for many non-equilibrium physical processes to be considered
within a general framework$^{11}$. \ Given the importance of this matter, it
is convenient to recall some essential aspects of the non-equilibrium theory,
leaving details to the dedicated treatises$^{12}$.

The basic equation of non-equilibrium thermodynamics is derived from the one
by Gibbs. In general, the entropy change of a system can be written as%
\begin{equation}
dS=d_{e}S+d_{i}S\label{21form1}%
\end{equation}
where $d_{e}S$ is due to interaction with the system surroundings (actually,
we are dealing with a closed system) and \ $d_{i}S$ \ is the entropy
production due to internal change of the system . \ In the case that internal
changes are due only to chemical reactions , the entropy production term can
be written as$^{11}$%

\begin{equation}
d_{i}S=-\frac{1}{T}\sum_{k=1}^{c}\mu_{k}dn_{k}\label{21form2}%
\end{equation}
where $\mu_{k}$ is the chemical potential of the k-component of a mixture
containing $c$ chemical species and $dn_{k}$ is the corresponding molar
change. For the ideal system $\mu_{k}=\zeta_{k}+RT\ln N_{k}$ where $\zeta_{k}$
is a quantity independent of actual composition and $N_{k}$ is the molar
fraction of the \ $k-$component. \ If the system holds $r$\ \ chemical
reactions, the \ molar change of the k-component can be written as
$dn_{k}=\sum_{l=1}^{r}\nu_{kl}d\xi_{l}$ \ where $\nu_{kl}$\ is the
stoichiometric coefficient of the k-component in the $l$-reaction which shows
a "displacement" $d\xi_{l}$ . \ Thus , the entropy production can also be
written as
\begin{equation}
d_{i}S=-\frac{1}{T}\sum_{kl}\mu_{k}\nu_{kl}d\xi_{l}=\frac{1}{T}\sum_{l}%
\Gamma_{l}d\xi_{l}\label{21form3}%
\end{equation}
where $\Gamma_{l}=$ -$\sum_{k}\mu_{k}\nu_{kl}=-\sum_{k}\nu_{kl}\zeta
_{k}-RT\sum_{k}\ln N_{k}^{\nu_{kl}}$ is the chemical affinity of the
$l$-reaction. At equilibrium the affinities vanish, so that \ $\sum_{k}%
\nu_{kl}\zeta_{k}=-RT\sum_{k}\ln\overline{N}_{k}^{\nu_{kl}}$. Thus, affinities
can also be written as \ $\Gamma_{l}=-RT\ln\Pi_{k}\left(  n_{k}/\overline
{n}_{k}\right)  ^{\nu_{kl}}$. \ If at any instant the entropy \ changes as a
function of chemical composition as well as other quantities characterizing
the system, it is possible to write an equation for the rate of entropy
production, that is, $dS/dt=d_{e}S/dt+d_{i}S/dt$, where%

\begin{equation}
\frac{d_{i}S}{dt}=\frac{1}{T}\sum_{l}\Gamma_{l}v_{l}\;,\label{21form4}%
\end{equation}
$v_{l}=d\xi_{l}/dt$ standing for the flux or velocity of the $l$-reaction.
\ It is assumed \ that near equilibrium fluxes are linear with respect
\ to\ the affinities, that is, $v_{l}=\sum_{m}L_{lm}\Gamma_{m}$ \ \ where
$L_{lm}$ are called the phenomenological coefficients$^{11}$ . As shown by
Onsager$^{13}$, \ based on the time reversal invariance of (microscopic)
mechanical laws, the phenomenological coefficients form a symmetric matrix,
that is $L_{lm}=L_{ml}$. It is to be pointed out that the criterion for sign
assignments to stoichiometric coefficients is quite arbitrary. However,
whatever the choice, if no external forces cause internal change,
$d_{i}S/dt\geqslant0$ holds.%
\[
\]

\begin{center}
\textbf{B. The variational calculus}\ 
\end{center}

The separation of entropy contributions given in eq. (\ref{21form1}) may cause
some problems of interpretation when the interaction with photons is
considered. \ Classically speaking, the interaction with the e.m. field
changes the density of internal energy with a rate given by the product
$\overrightarrow{E}\,\overrightarrow{j}$ where $\overrightarrow{E}\,$\ stand
for the strength of electric field and $\overrightarrow{j}$ \ for the vector
of current density. Thus, a field releasing its energy to (emitted from) the
system increases (decreases) the internal energy and thus the entropy. If
configurational or chemical changes of the system are involved, the picture is
slightly more complex. Actually, the e.m. field, which causes departure of the
chemical reactions from equilibrium, reduces the entropy. A sequence of
reactions brings the system to equilibrium, thus producing positive entropy.
At the end of the process, the net change of chemical entropy is null. But
reactions may produce heat and photons which alter the internal energy and
thus the system entropy. Thus chemical reactions behave like a machine
converting the absorbed photons to heat (and photons with lesser energies).
About the entropy balance, further considerations will be advanced in the Sec.
IV. Now it appears convenient to write the rate of internal entropy
production\ $d_{i}S/dt$ \ as the sum of two contributions, that is,%

\begin{equation}
\frac{d_{i}S}{dt}=\sigma_{ext}+\sigma_{int}\label{21formYY1}%
\end{equation}
where $\sigma_{ext}$ means the change of internal entropy due to external
force and $\sigma_{int}$ that due to the internal \ forces, that is, those
bringing the system to equilibrium. The definitions of these two terms are to
be searched for by means of the composition changes \ induced by absorbed
photons, that is,%

\begin{equation}
dn_{k}=c_{k}\Phi dt+\sum_{l=1}^{r}\nu_{kl}d\xi_{l}\label{21formYY2}%
\end{equation}
where $\Phi$ stands for the flux of the whole absorbed photons (in suitable
units) and $c_{k}$ for the fraction of photon flux inducing molar change of
the k-component. \ It follows from eqs (\ref{21form3}) and (\ref{21formYY2}) that\ %

\begin{equation}
\frac{d_{i}S}{dt}=-\frac{1}{T}\Phi\Gamma_{\Phi}+\frac{1}{T}\sum_{l}\Gamma
_{l}v_{l}\label{21formXX1}%
\end{equation}
where $\Gamma_{\Phi}=\sum_{k}c_{k}\mu_{k}$ \ in the following will be referred
to as external affinity or "force"$^{11}$. \ We define%

\begin{equation}
\sigma_{ext}=-\frac{1}{T}\Phi\Gamma_{\Phi}\qquad\sigma_{int}=\frac{1}{T}%
\sum_{l}\Gamma_{l}v_{l}\;.\label{21formYY3}%
\end{equation}
The equation (\ref{21formXX1}) can be generalized to include more independent
sets of chemical reactions by adding further index labelling reaction sets,
that is,%
\begin{equation}
\sigma_{int}=\frac{1}{T}\sum_{\gamma l}\Gamma_{\gamma l}v_{\gamma
l}\;.\label{21formYY4}%
\end{equation}

Variational procedures allowing for the MREP must account for the constraints
that keep the system from equilibrium. These can be formalized by means of
relations among affinities which are to be inserted in a Lagrange minimization
procedure, where the affinities \ are the functional variables. \ We will see
in Sec. 3 that the constraints have the form%

\begin{equation}
\Gamma_{\Phi}-\sum_{j}\Gamma_{\gamma j}=0\qquad\gamma
=1,...,r\;.\label{21formYY19}%
\end{equation}
It is to be remarked that minimization concerns only $\sigma_{int}$\ , that
is, only the internal processes bringing the system to equilibrium. However,
to get a complete definition of the steady state, it is convenient to apply
the minimization procedure to the function
\begin{equation}
\Omega=2\sigma_{ext}+\sigma_{int}+\frac{1}{T}\sum_{\gamma}\lambda_{\gamma
}\left(  \Gamma_{\Phi}-\sum_{j}\Gamma_{\gamma j}\right)  \;,\label{21formYY6}%
\end{equation}
by including $\Gamma_{\Phi}$ among the functional variables. Factor 2 takes
into account that $\sigma_{int}$ is a quadratic form on the affinities(this
procedure is substantially different from that used in ref. [10]). As a
result, it is obtained that $\Phi=\sum_{\gamma}\lambda_{\gamma}/2$ \ and
$\upsilon_{\gamma}\equiv\upsilon_{\gamma l}=\lambda_{\gamma}/2$ \ $l=1,...,r$
, that is, $\Phi=\sum_{\gamma}\upsilon_{\gamma}$. Thus, at the steady state
the fluxes within a reaction set are the same for all reactions and the whole
\ flux of reaction sets is equal to the flux of absorbed photons. \ 

In this paper we are concerned with reactions allowing for a diagonal
Onsager's matrix (see also appendix A). The diagonal Onsager's matrix is
peculiar to reactions that do not show interference effects so that$^{11}$%
\begin{equation}
\upsilon_{\gamma l}=L_{\gamma l}\Gamma_{\gamma l}\qquad
l=1,...,c\;.\label{21formYY7}%
\end{equation}
Thus, by taking into account eq. (\ref{21formYY19}) it follows
\begin{equation}
\upsilon_{\gamma}=\frac{\Gamma_{\Phi}}{\sum_{j}1/L_{\gamma j}}%
\;.\label{21formYY9}%
\end{equation}
Note that, at the steady state, flux is independent of the reaction index but
depends only on the reaction-set index. The reaction affinity is related to
the external force by the equation
\begin{equation}
\Gamma_{\gamma l}=\Gamma_{\Phi}\frac{1/L_{\gamma l}}{\sum_{j}1/L_{\gamma j}%
}\;.\label{21formYY10}%
\end{equation}
Now, by defining the RRs as%

\begin{equation}
R_{\gamma l}=1/L_{\gamma l}\;,\label{21formYY11}%
\end{equation}
we are able to describe a set of chemical reactions, activated by the photon
flux with external force \ $\Gamma_{\Phi}$ , as a current $\ \upsilon_{\gamma
}$ passing through a series of resistances $R_{\gamma j}$ to which is applied
a voltage $\Gamma_{\Phi}.$ Thus, the drop in voltage \ on the resistance
$R_{\gamma j}$ , that is, $\Gamma_{\gamma l}=$\ $R_{\gamma j}\upsilon_{\gamma
}$, can be calculated by means of eq. (\ref{21formYY10}) which gives the
voltage partition, that is,%
\begin{equation}
\Gamma_{\gamma l}=\frac{R_{\gamma l}}{\sum_{j}R_{\gamma j}}\Gamma_{\Phi
}\;.\label{21formYY12}%
\end{equation}
Finally, for full correspondence to Ohm's law we must give a meaning to the
Joule equation for the dissipated power, that is,
\begin{equation}
W_{\gamma}=\Gamma_{\Phi}\upsilon_{\gamma}=\upsilon_{\gamma}\sum_{j}%
\Gamma_{\gamma j}=T\;\sigma_{int}\label{21formYY13}%
\end{equation}
which corresponds to the rate of entropy production of the $\gamma$-reaction
set \ multiplied by temperature. \ This has an evident meaning: as the
electrical work dissipate the electric potential energy by producing heat, the
chemical reactions dissipate the stored chemical energy by producing entropy
(in general there is production of heat and photons).

Variational calculus can be addressed to obtain information about the
distribution of reaction fluxes. To this end we should minimize \ $\sigma
_{int}$ \ with respect to the fluxes by putting \ $\upsilon_{\gamma
l}=\upsilon_{\gamma}$ \ $l=1,2,...,c$, $\Gamma_{\gamma l}=\upsilon_{\gamma
l}/L_{\gamma l}=\upsilon_{\gamma}/L_{\gamma l}$\ \ $\gamma=1,2,...,r$ and
\ $\Phi-\sum_{\gamma}\upsilon_{\gamma}=0$. Minimization is to be applied to
the function%

\begin{equation}
\Omega^{\ast}=2\sigma_{ext}+\ \sigma_{int}+\frac{\chi}{T}\left(  \Phi
-\sum_{\gamma}\upsilon_{\gamma}\right)  \;,\label{21formYY14}%
\end{equation}
that is, explicitly,%

\begin{equation}
\Omega^{\ast}=-\frac{2}{T}\Phi\Gamma_{\Phi}+\frac{1}{T}\sum_{\gamma}%
\upsilon_{\gamma}^{2}\sum_{l}\frac{1}{L_{\gamma l}}+\frac{\chi}{T}\left(
\Phi-\sum_{\gamma}\upsilon_{\gamma}\right) \label{21formYY15}%
\end{equation}
where $\Phi$ is to be included among the functional variables. By defining the
equivalent phenomenological coefficient of the $\gamma$-reaction set as%

\begin{equation}
L_{\gamma EQ}=1/\sum_{l}\frac{1}{L_{\gamma l}}\label{21formYY16}%
\end{equation}
it is obtained from the variational calculus that \ $\Gamma_{\Phi}=\chi/2$ and
$\upsilon_{\gamma}=L_{\gamma EQ}\chi/2$, that is, \
\begin{equation}
\upsilon_{\gamma}=L_{\gamma EQ}\Gamma_{\Phi}=\Phi L_{\gamma EQ}/\sum_{\gamma
}L_{\gamma EQ}\label{21formYY17}%
\end{equation}
which is the equation of the partition current provided we define the
equivalent resistance of the $\gamma$-reaction set as (see eqs.
\ref{21formYY11} and \ref{21formYY17} )%

\begin{equation}
R_{\gamma EQ}=1/L_{\gamma EQ}=\sum_{j=1}^{c}R_{\gamma j}\;.\label{21formYY18}%
\end{equation}%
\[
\]

\begin{center}
{\large III. APPLICATION TO ELECTRON-HOLE RECOMBINATIONS}%
\[
\]

\end{center}

Until now, for simplicity, we have considered cases where internal changes are
to be ascribed to chemical reactions. To deal with electron-hole
recombinations we should consider more suitable units , that is, densities
($cm^{-3}$) rather than molar concentrations. To this end, we must divide the
rate of entropy production by molar volume, $V_{M}$ ($cm^{3}$), so that%

\[
\frac{\upsilon\Gamma}{V_{M}}=-\frac{d\left(  N_{0}\xi/V_{M}\right)  }{dt}%
kT\ln\Pi_{k}\left(  n_{k}/\overline{n}_{k}\right)  ^{\nu_{k}}
\]
where $N_{0}$ \ stands for the Avogadro number, the other symbols having the
usual meanings. Now, $dN_{0}\xi/V_{M}/dt$ is the reaction flux with the
desired units, that is, $cm^{-3}s^{-1}$ . In the following, as a consequence
of this unit choice, the affinities will be calculated as%

\[
\Gamma=-kT\ln\Pi_{k}\left(  n_{k}/\overline{n}_{k}\right)  ^{\nu_{k}}\;.
\]%
\[
\]

\begin{center}
\textbf{A. The SK and LK cases}
\end{center}

Near equilibrium (but also in most practical cases) the rate equations can be
written in the approximation of non-degenerate statistics, that is, emission
rates of trapped carriers released into bands are independent of the
occupation of band levels. In this approximation, rate equations assume a very
simple form$^{8,10}$. For simplicity's sake, it is convenient to begin by
considering the cases of SK and LK . Formally, they can be dealt with as a
single case. Indeed, they differ only in what kind of carrier is trapped and
what is recombined, that is, if an electron is recombined with a trapped hole
or, conversely, if a hole is recombined with a trapped electron. As stated in
Section I, this difference has no formal relevance since it is due only to
actual level position with respect to the equilibrium Fermi level

Let $N_{t}$ mean the density of defects, $p_{t}$ the actual density of trapped
holes (traps empty of electrons), $n_{t}$ the density of traps empty of holes
( filled with electrons), so that $N_{t}=p_{t}+n_{t}$. \ Owing to interactions
with ionizing radiations, electrons and holes are injected into conduction and
valence bands respectively, with a rate $\Phi$. Carriers are captured with
probability $c_{e}$, for conduction band electrons, and $c_{p}$ , for valence
band holes, respectively. Thermal releasing of electrons into conduction band
and of holes into valence band occurs with probabilities $s_{e}$ and $s_{p}%
$\ (included is the interaction with black-body radiation), respectively.
Thus, the net rate of electron capture is$^{8}$
\begin{equation}
\upsilon_{n}=c_{e}np_{t}-s_{e}n_{t}\label{Form9}%
\end{equation}
and that of hole capture is%

\begin{equation}
\upsilon_{p}=s_{p}p_{t}-c_{p}pn_{t}\label{Form10}%
\end{equation}
At equilibrium $\ \upsilon_{n}=\upsilon_{p}=0$. Thus,%

\begin{equation}
\frac{\overline{n}\overline{p}_{t}}{\overline{n}_{t}}=\frac{s_{e}}{c_{e}%
}\label{21formSK1}%
\end{equation}

\begin{equation}
\frac{\overline{p}\overline{n}_{t}}{\overline{p}_{t}}=\frac{s_{p}}{c_{p}%
}\label{21formSK2}%
\end{equation}
To define affinities, we need to fix a positive direction for processes as a
sign reference for stoichiometric coefficients. Let us take as the positive
direction that of the arrow pointing from the conduction to the valence band.
Thus, the terms representing processes which bring electrons towards the
valence band, as well as holes towards the conduction band, are associated
with the stoichiometric coefficient equal to +1. The opposite sign is
associated with the terms describing processes in the opposite direction. On
this ground, the electron affinity is
\begin{equation}
\Gamma_{n}=-kT\ln\left[  \left(  \frac{n}{\overline{n}}\right)  ^{\nu_{a}%
}\left(  \frac{n_{t}}{\overline{n}_{t}}\right)  ^{\nu_{b}}\left(  \frac{p_{t}%
}{\overline{p}_{t}}\right)  ^{\nu_{c}}\right] \label{21formSK3}%
\end{equation}
with $\nu_{a}=\nu_{c}=+1$, since $n$ and $p_{t}$ appear in a term of positive
direction, and $\nu_{b}=-1$, since $n_{t}$ appears in a term of negative
direction. Thus, we can write%

\begin{equation}
\Gamma_{n}=kT\ln\left(  \frac{n_{t}}{np_{t}}\frac{\overline{n}\overline{p}%
_{t}}{\overline{n}_{t}}\right)  \;.\label{21formSK4}%
\end{equation}
The hole affinity can be obtained in an analogous way, that is,%

\begin{equation}
\Gamma_{p}=kT\ln\left(  \frac{p_{t}}{pn_{t}}\frac{\overline{p}\overline{n}%
_{t}}{\overline{p}_{t}}\right)  \;.\label{21formSK5}%
\end{equation}
The external affinity is%
\[
\Gamma_{\Phi}=-kT\ln\left[  \left(  \frac{n}{\overline{n}}\right)  ^{c_{a}%
}\left(  \frac{p}{\overline{p}}\right)  ^{c_{b}}\right]
\]
with c$_{a}=c_{b}=-1$ since the reaction is opposite to the fixed positive
direction. Thus%

\begin{equation}
\Gamma_{\Phi}=kT\ln\left(  \frac{np}{\overline{n}\overline{p}}\right)
\;,\label{21formSK6}%
\end{equation}
so that the loop constraint (\ref{21formYY19}) is satisfied, that is,
$\Gamma_{\Phi}-\left(  \Gamma_{1}+\Gamma_{2}\right)  =0$. For steady states
near equilibrium we can be write, approximately,%

\begin{equation}
\upsilon_{n}=c_{e}\overline{n}\overline{p}_{t}\left(  1-\frac{n_{t}}{np_{t}%
}\frac{\overline{n}\overline{p}_{t}}{\overline{n}_{t}}\right)
\;,\label{21formSK7}%
\end{equation}%
\begin{equation}
\upsilon_{p}=c_{p}\overline{p}\overline{n}_{t}\left(  1-\frac{p}{pn_{t}}%
\frac{\overline{p}\overline{n}_{t}}{\overline{p}_{t}}\right)
\;.\label{21formSK8}%
\end{equation}
In the same approximation, the latter eqs. can be re-written as%

\begin{equation}
\upsilon_{n}=\frac{c_{e}\overline{n}\overline{p}_{t}}{kT}\Gamma_{n}%
=L_{n}\Gamma_{n}\;,\label{21formSK9}%
\end{equation}

\begin{equation}
\upsilon_{p}=\frac{c_{p}\overline{p}\overline{n}_{t}}{kT}\Gamma_{p}%
=L_{p}\Gamma_{p}\;.\label{21formSK10}%
\end{equation}
where%
\begin{equation}
L_{n}=\frac{c_{e}\overline{n}\overline{p}_{t}}{kT}\qquad L_{p}=\frac
{c_{p}\overline{p}\overline{n}_{t}}{kT}\label{21formSK11}%
\end{equation}
are the Onsager's coefficients. Finally, by definitions, the RRs are%
\[
R_{n}=\frac{1}{L_{n}}=\frac{kT}{c_{e}\overline{n}\overline{p}_{t}}\;,
\]%
\[
R_{p}=\frac{1}{L_{p}}=\frac{kT}{c_{p}\overline{p}\overline{n}_{t}}\;,
\]
which agree with that found by SR (see eqs. (\ref{IntroX1}) and (\ref{IntroX2}%
)). Note that the steady current is%

\[
\upsilon=\upsilon_{n}=\upsilon_{p}=\frac{\Gamma_{\Phi}}{R_{n}+R_{p}}\;,
\]
so that%

\[
\upsilon\left(  R_{n}+R_{p}\right)  =\Gamma_{\Phi}\;.
\]
By taking into account the definitions of QFLs for free carriers (see
eq.\ref{IntroZZ1}), it follows,
\[
\Gamma_{\Phi}=kT\ln np/\overline{n}\overline{p}=F_{n}-F_{p}\;,
\]
so that%

\[
\upsilon\left(  R_{n}+R_{p}\right)  =F_{n}-F_{p}\;,
\]
which is exactly the result found by SR (eq. \ref{IntroX3}).

Note also that , according to SR (by integrating the eq. 2.9 of ref. [8] over
the whole states of conduction band)%

\begin{equation}
\frac{\overline{n}\overline{p}_{t}}{\overline{n}_{t}}=\frac{s_{e}}{c_{e}%
}=N_{n}\exp\left[  \left(  E_{t}-E_{C}\right)  /kT\right] \label{21AGGYY2}%
\end{equation}
where $E_{t}$ stands for the energy of trapping level. Now, by taking into
account that%

\begin{equation}
\frac{n_{t}}{p_{t}}=\frac{f_{t}}{1-f_{t}}=\exp(E_{t}-F_{t})\label{21AGGYY3}%
\end{equation}
$f_{t\text{ }}$ being the occupancy of electron traps with QFL $F_{t}$, it
follows from eqs \ (\ref{21formSK4}) and (\ref{21formSK5}) that
\begin{equation}
\Gamma_{n}=F_{n}-F_{t}\;.\label{21AGGXX3}%
\end{equation}
Analogously, for the affinity of trapped hole it can be shown that%
\begin{equation}
\Gamma_{p}=F_{t}-F_{p}\;.\label{21AGGXX4}%
\end{equation}
The eqs (\ref{21AGGXX3}) and (\ref{21AGGXX4}) allow for a full correspondence
to the SR results.%
\[
\]

\begin{center}
\textbf{B. The PW case}
\end{center}

SR in their paper considered only recombinations in single level centres, as
in SK or LK processes. Now, it is advisable to consider also the case of
two-level recombination centres as in the PW processes. With respect to the SK
(or LK) two-step processes, in the PW case we must consider a further step,
that is, the recombinations of electrons trapped in levels labelled, say, 1
with holes trapped in levels labelled, say, 2. \ The corresponding
recombination rate is
\begin{equation}
\upsilon_{\omega}=\pi n_{1t}p_{2t}-s_{\omega}p_{1t}n_{2t}\label{21formPW1}%
\end{equation}
where $\pi$ stands for the recombination probability, $n_{1t}$\ and $p_{1t}%
$\ for the densities of electrons and holes trapped at the level labelled 1
respectively, $n_{1t}$\ and $p_{1t}$\ for the densities of electrons and holes
trapped at the level labelled 2 respectively and $s_{\omega}$ for the
probability of pair production by thermal excitation. At equilibrium%

\begin{equation}
\frac{\overline{n}_{1t}\overline{p}_{2t}}{\overline{p}_{1t}\overline{n}_{2t}%
}=\frac{s_{\omega}}{p}\;,\label{21formPW2}%
\end{equation}
so that%
\begin{equation}
\upsilon_{\omega}=\pi\overline{n}_{1t}\overline{p}_{2t}\left(  1-\frac
{p_{1t}n_{2t}}{n_{1t}p_{2t}}\frac{\overline{n}_{1t}\overline{p}_{2t}%
}{\overline{p}_{1t}\overline{n}_{2t}}\right)  \;.\label{21formPW3}%
\end{equation}
It is easy to verify that the affinity is%

\begin{equation}
\Gamma_{\omega}=kT\ln\frac{p_{1t}n_{2t}}{n_{1t}p_{2t}}\frac{\overline{n}%
_{1t}\overline{p}_{2t}}{\overline{p}_{1t}\overline{n}_{2t}}%
\;.\label{21formPWW}%
\end{equation}
Thus by using the definitions of electron and hole affinities (obtained in the
previous section) suitably modified to account for the index levels , it is
easy to prove that $\Gamma_{\Phi}=\Gamma_{n1}+\Gamma_{\omega}+\Gamma_{p2}\;$.
Now, the flux of the recombination process is%
\begin{equation}
\upsilon_{\omega}=L_{\omega}\Gamma_{\omega}=L_{\omega}\left[  \Gamma_{\Phi
}-\left(  \Gamma_{n1}+\Gamma_{p2}\right)  \right] \label{21formPW4}%
\end{equation}
where
\begin{equation}
L_{\omega}=\pi\overline{n}_{1t}\overline{p}_{2t}/kT\label{21formPW5}%
\end{equation}
to which is associated the RR
\begin{equation}
R_{\omega}=1/L_{\omega}=kT/\pi\overline{n}_{1t}\overline{p}_{2t}%
\;.\label{21formPW6}%
\end{equation}
The current through the recombination channel is%

\begin{equation}
\upsilon=\frac{\Gamma_{\Phi}}{R_{n1}+R_{p2}+R_{\omega}}\;.\label{21formPW7}%
\end{equation}
Note that the affinity $\Gamma_{\omega}$ is related to QFLs by the equation
$\Gamma_{\omega}=F_{t1}-F_{t2}$.$\;$
\[
\]

\begin{center}
\textbf{C. The case of band-to-band recombination}
\end{center}

As a final example, let us calculate the RR associated with band-to-band
recombinations which now appears as an easy task. Briefly, the net rate of
recombinations is $\upsilon_{G}=\pi_{G}np-s_{G}$ \ where \ $\pi_{G}$\ stands
for the probability of electron-hole pair recombination and $s_{G}$ for the
thermal emission of electrons from the valence to conduction band. At
equilibrium $\overline{n}\overline{p}=s_{G}/\pi_{G}$. Thus $\upsilon_{G}%
=\pi_{G}\overline{n}\overline{p}\left(  1-\overline{n}\overline{p}/np\right)
=L_{G}\Gamma_{G}$ where $\Gamma_{G}=\Gamma_{\Phi}$ and $L_{G}=\pi_{G}%
\overline{n}\overline{p}/kT$. $\;$The associated RR is $R_{G}=1/L_{G}%
=kT/\pi_{G}\overline{n}\overline{p}$, $\;$so that the current can be written
as $\upsilon_{G}=\Gamma_{\Phi}/R_{G}$.
\[
\]
$\;.$

\begin{center}
{\large IV. FINAL REMARKS AND CONCLUSIONS}%
\[
\]

\end{center}

At the steady state, the entropy \ of the system is time independent so that
$dS/dt=0$ , that is, $d_{e}S/dt=-d_{i}S/dt$ . From eq. (\ref{21formXX1}) \ and
by taking into account of the results of variational calculus we see that
$d_{i}S/dt=0$. This result is expected since the chemical composition is time
independent as well. Nevertheless, it is worth dwelling upon this point to
complete the considerations in Section II. Actually, the entropy balance is
based on the following scheme:%

\[
photons~(h\nu)\longrightarrow Chemical~reactions\longrightarrow
heat+photons~(h\nu^{\prime}\leq h\nu)
\]
A flux of energy carried by photons, owing to internal reactions, is partly
converted \ to heat and partly to photons (of lesser energy). \ The rate of
heat production is thus $dQ/dt=\Phi\left(  h\nu-h\nu^{\prime}\right)  $ which,
divided by the system temperature, is to be considered as the indirect
contribution to $d_{e}S/dt$ due to the flux of ionizing photons. But the
system is not isolated, so that the heat produced by reactions is thus
released to surroundings. This causes at the steady state $d_{e}S/dt=0$ but
\ a positive entropy is released to the surroundings thus contributing to the
increase in the entropy of the universe.

The definitions of RR rely on the diagonal property of the Onsager's matrix
associated with the recombination processes. Besides some nice cases, the
generalization to non-diagonal processes does not appear to be possible. In
Appendix B we try to diagonalize the thermodynamic problem of a two-reaction
process that satisfies eq. (\ref{21formYY11}).

We again stress that the results found in this paper hold in the linear range
of non-equilibrium thermodynamics. In this connection, it is understood that
they are to be considered in the same spirit as one considers the linear
approximation when small deviations from linearity can be disregarded$^{14}$.
\ The validity of MREP far from equilibrium is still debated$^{15}$.
\ Moreover, in general, relations among fluxes and forces are not linear.
However, linearization of relations is possible provided the system dealt with
is near the steady state considered as the "reference state"$^{16}$.
\ Unfortunately, the matrix of coefficients is not symmetric for long. Thus,
at the moment, definitions of RRs (which conserve thermodynamic meanings) for
processes far from equilibrium appear a difficult task.

In this paper the electron-hole recombinations \ in a excited material near
equilibrium has been investigated within the framework of the non-equilibrium
thermodynamics ruled by the MREP principle. A formal Ohm-like law has been
proved to hold for recombinations, thus leading in a natural way to the
concept of recombination resistances, as defined by Shokley and Read. The
entropy produced by recombinations is found to obey a Joule-like law where it
(multiplied by temperature) play the role of the "dissipated power". This
makes the correspondence between steady electron-hole recombinations and
electrical processes complete, at least when linear flux laws hold.%
\[
\]

\begin{center}
{\large APPENDIX A}%
\[
\]

\end{center}

The reactions dealt with in this paper can be represented by means of a
chain-process scheme, that is,%

\[
n_{1}\rightleftarrows n_{2}\rightleftarrows...\rightleftarrows n_{l}%
\rightleftarrows...\rightleftarrows n_{c}
\]
where the $l-th$ reactions depend only on the $l$ and the $l+1$ components.
This picture leads to a simple form for affinities, that is,
\[
\Gamma_{l}=\nu_{l,l+1}\mu_{l}+\nu_{l+1,l}\mu_{l+1}
\]
with $\nu_{l,l+1}=-\nu_{l+1,l}=1$ . \ By definition it follows that
$\upsilon_{l}=L_{l}\Gamma_{l}$ thus giving a diagonal Onsager's matrix. Now,
let us suppose that the set of reactions form a closed loop, that is,%

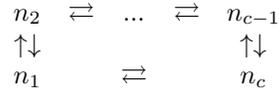
\[%
\begin{array}
[c]{ccccc}%
n_{2} & \rightleftarrows & ... & \rightleftarrows & n_{c-1}\\
\uparrow\downarrow &  &  &  & \uparrow\downarrow\\
n_{1} &  & \rightleftarrows &  & n_{c}%
\end{array}
\]
In this case $\sum_{l}\Gamma_{l}=0$, that is
\[
\sum_{l\neq c}\Gamma_{l}=-\Gamma_{c}=-\left(  \nu_{c,1}\mu_{c}+\nu_{1,c}%
\mu_{1}\right)  =\mu_{1}-\mu_{c}
\]
If we consider an external force that moves the system from equilibrium in the
point between components 1 and c, then it is opposite to the $\Gamma_{c} $
force that leads the system towards equilibrium. Thus, by labelling this force
with $\Phi$ we can write%

\[
\Gamma_{\Phi}=-\Gamma_{c}=\sum_{l\neq c}\Gamma_{l}
\]
from which we obtain eq. (\ref{21formYY19}).

\begin{center}%
\[
\]
{\large APPENDIX B}%
\[
\]

\end{center}

Now let us consider a case where eq. (\ref{21formYY19}) holds but the
symmetric Onsager's matrix is not diagonal. Let us consider a two-component
reaction set with fluxes%
\begin{align*}
\upsilon_{1}  & =L_{11}\Gamma_{1}+L_{12}\Gamma_{2}\\
\upsilon_{2}  & =L_{21}\Gamma_{1}+L_{22}\Gamma_{2}%
\end{align*}
which can be re-written as%

\begin{align*}
\upsilon_{1}  & =\left(  L_{11}-L_{12}\right)  \Gamma_{1}+L_{12}\Gamma_{\Phi
}\\
\upsilon_{2}  & =L_{21}\Gamma_{\Phi}+\left(  L_{22}-L_{21}\right)  \Gamma_{2}%
\end{align*}
At the steady state $\upsilon_{1}=\upsilon_{2}$ so that%

\[
\Gamma_{i}=\frac{\left(  L_{jj}-\eta\right)  }{\left(  L_{11}-\eta\right)
+\left(  L_{22}-\eta\right)  }\Gamma_{\Phi}
\]
where j$\neq i$ and $\eta=L_{12}=L_{21}$. Moreover we see that
\[
\upsilon_{i}=\left[  \frac{\left(  L_{11}-\eta\right)  \left(  L_{22}%
-\eta\right)  }{\left(  L_{11}-\eta\right)  +\left(  L_{22}-\eta\right)
}+\eta\right]  \Gamma_{\Phi}
\]
which does not has the formal structure of Ohm's law. But, by defining the new fluxes%

\[
\widetilde{\upsilon}_{i}=\upsilon_{i}-\eta\Gamma_{\Phi}=\frac{\left(
L_{11}-\eta\right)  \left(  L_{22}-\eta\right)  }{\left(  L_{11}-\eta\right)
+\left(  L_{22}-\eta\right)  }\Gamma_{\Phi}
\]
and \ the new phenomenological\ coefficients%

\[
\widetilde{L}_{ii}=L_{ii}-\eta
\]
we are able to diagonalize the thermodynamic problem provided a new flux is
considered, that is,%

\[
\widetilde{\upsilon}_{\Phi}=\eta\Gamma_{\Phi}
\]
Consequently, the variational calculus is to be applied to the function%

\[
\Omega=-2\frac{\Phi\Gamma_{\Phi}}{T}+\frac{\widetilde{\upsilon}_{1}\Gamma_{1}%
}{T}+\frac{\widetilde{\upsilon}_{2}\Gamma_{2}}{T}+\frac{\widetilde{\upsilon
}_{\Phi}\Gamma_{\Phi}}{T}+\frac{1}{T}\lambda\left[  \Gamma_{\Phi}-\left(
\Gamma_{1}+\Gamma_{2}\right)  \right]
\]
Now, we can define the reaction resistances as $R_{i}=1/\left(  L_{ii}%
-\eta\right)  $ \ and $R_{\Phi}=1/\eta$ , where $R_{\Phi}$ is to be considered
in parallel with the other two resistances. Note that in the case L$_{ii}%
=\eta$ it follows that \ $\upsilon_{1}=\upsilon_{2}=\eta\Gamma_{\Phi}$, and
$\widetilde{\upsilon}_{1}=\widetilde{\upsilon}_{2}=0$ which is consistent with
the infinite resistance.%
\[
\]
(*) Electronic address: masalis@vaxca1.unica.it%

\[
\]
$^{1}$C. Kittel, Introduction to Solid State Physics (John Wiley \& Sons, Inc.
New York, London, Sydney 1966).%

\[
\]
$^{2}$D. Curie, Luminescence in Crystals (Methuen, London 1963).%

\[
\]
$^{3}$A. S$.$ Grove, Physics and Technology of Semiconductor Devices (John
Wiley and Sons, Inc., New York, London, Sydney, 1967)%

\[
\]
$^{4}$C. P. Flynn, Point Defects and Diffusion (Clarendon Press, Oxford 1972).%

\[
\]
$^{5}$J. C. Simmons and G. W. Taylor, Phys. Rev. B \textbf{4} 502 (1971).%

\[
\]
$^{6}$R.Chen and S. W. S. McKeever, Theory of Thermoluminescence and Related
Phenomena (World Scientific, \ Singapore, \ New Jersey, London, Hong Kong, \ 1997).%

\[
\]
$^{7}$See for example: H. M. Chen, Y. F. Chen, M. C. Lee, M. S. Feng, Phys.
Rev. B, \textbf{56}, 6942 (1997)%

\[
\]
$^{8}$W. Shockley and \ W. T. Read, Jr. , Phys. Rev. \textbf{87} 835 (1952).%

\[
\]
$^{9}$D. Vanmaekelbergh and F. Cardon, Semicon. Sci Technol. \textbf{3}, 124 (1988).%

\[
\]
$^{10}$V. Maxia, Phys. Rev. B \textbf{25}, 4196 (1982); Phys. Rev. B,
\textbf{21} 749 (1980); Phys. Rev. B \textbf{17}, 3262 (1978).%

\[
\]
$^{11.}$I$.$Prigogine, Introduction to Thermodynamics of Irreversible
Processes , (John Wiley\&Sons, Interscience, N. Y., 1971).%

\[
\]
$^{12}$H.J. Kreuzer, Nonequilibrium Thermodynamics and its Statistical
Foundations, (Clarendon press, Oxford 1981).%

\[
\]
$^{13}$L. Onsager, Phys. Rev., \textbf{37},405 (1931); \textbf{38},2265 (1931).%

\[
\]
$^{14}$ D. K. Kondepudi, Physica A \textbf{154}, 204 (1988)%

\[
\]
$^{15}$See for example: R. J. Tykodi, Physica \textbf{72, }341 (1974); D. J.
Evans and A. Baranyai, Phys. Rev. Lett. \textbf{67}, 2597 (1991); A. R.
Vasconcellos and R. Luzzi Phys. A. \textbf{180}, 182 (1992).%

\[
\]
$^{16}$G. Nicolis and I. Prigogine, Self-Organization in Nonequilibrium
Systems (John Wiley \& Sons New York, London, Sydney, Toronto, 1977).

\end{document}